\documentclass{article}

\usepackage[numbers]{natbib}
\usepackage[hyphens]{url}

\usepackage[final]{neurips_2024}

\usepackage[utf8]{inputenc} 
\usepackage[T1]{fontenc}    
\usepackage{hyperref}       
\usepackage{url}            
\usepackage{booktabs}       
\usepackage{amsfonts}       
\usepackage{nicefrac}       
\usepackage{microtype}      
\usepackage{xcolor}         

\usepackage{amsmath}
\usepackage{mathtools}
\usepackage{amsthm}
\usepackage{scrextend}
\usepackage{graphicx}
\usepackage{amssymb}
\usepackage{algpseudocode}
\usepackage{algorithm}
\usepackage{booktabs}
\usepackage[subrefformat=parens]{subcaption}
\floatstyle{plaintop}
\restylefloat{table}

\title{FlexFlood: Efficiently Updatable \\ Learned Multi-dimensional Index}

\author{%
  Fuma Hidaka\\
  The University of Tokyo\\
  \texttt{hidaka@hal.t.u-tokyo.ac.jp} \\
  \And
  Yusuke Matsui\\
  The University of Tokyo\\
  \texttt{matsui@hal.t.u-tokyo.ac.jp} \\
}

\begin{document}

\maketitle

\begin{abstract}
  A learned multi-dimensional index is a data structure that efficiently answers multi-dimensional orthogonal queries by understanding the data distribution using machine learning models. One of the existing problems is that the search performance significantly decreases when the distribution of data stored in the data structure becomes skewed due to update operations. To overcome this problem, we propose FlexFlood, a flexible variant of Flood. FlexFlood partially reconstructs the internal structure when the data distribution becomes skewed. Moreover, FlexFlood is the first learned multi-dimensional index that guarantees the time complexity of the update operation. Through experiments using both artificial and real-world data, we demonstrate that the search performance when the data distribution becomes skewed is up to 10 times faster than existing methods. We also found that partial reconstruction takes only about twice as much time as naive data updating.
\end{abstract}

\section{Introduction}\label{sec:Introduction}
Filtering, scanning, and updating of data are fundamental operations for databases, and various data structures have been studied to perform these operations efficiently. In the real world, we often need to handle multi-dimensional data, and Kd-tree and its variants~\cite{KdTree, RTree, ZOrderCurve} are typical data structures for handling them. These data structures are widely used in real-world applications~\cite{Redshift, SparkSQL, IBMInformix}.

Recently, there has been active research to improve data structures by learning the distribution of data and queries with machine learning models. Such data structures are called learned index~\cite{LearnedIndex}. One of the significant challenges in learned multi-dimensional indexes~\cite{Flood, Tsunami, LISA, RLRTree, Waffle} is that many do not support data update operations. Even if they do, none describe the time complexity for updating.

Therefore, we proposed a flexible variant of Flood (FlexFlood) that supports efficient data updating by adaptively modifying the internal structure of the existing learned multi-dimensional index, Flood~\cite{Flood}. We proved that the amortized time complexity of updating is $O(D \log N)$ under two assumptions that the data increases at an approximately constant pace and that the training results of the ML model satisfy certain conditions. Here, $D$ is the dimensionality of the data, and $N$ is the total number of data. Furthermore, experiments using multiple artificial and real-world datasets confirm that the ``certain conditions'' can be sufficient. The original Flood's search speed slows down as we update data, but FlexFlood remains fast. As a result, FlexFlood is up to 10 times faster than Flood after many updates. The source code for FlexFlood is available at \href{https://github.com/mti-lab/FlexFlood}{https://github.com/mti-lab/FlexFlood}.

\section{Related Work}\label{sec:RelatedWork}
Many classical data structures have been proposed for handling multi-dimensional data. For example, tree-based structures include Kd-tree~\cite{KdTree, ScapegoatTrees}, R-tree~\cite{RTree, R*Tree}, and Oct-tree~\cite{OctTree}. Grid File~\cite{GridFile} is also proposed as a space-partitioning structure. Additionally, there is the Z-order curve algorithm~\cite{ZOrderCurve}, which reduces multi-dimensional data into one dimension using a special sorting technique.

Learned index~\cite{LearnedIndex, LearnedDataStructure, LearnedIndexSurvey} is a data structure that incorporates machine learning models into classical data structures such as B-tree~\cite{BTree}, Hash Map~\cite{HashMap}, and Bloom Filter~\cite{BloomFilter}. Learned indexes improve performance by taking advantage of the distribution of data and queries. Recently, learned indexes have been actively researched. For example, several learned Bloom Filters have been proposed~\cite{LearnedBloomFilterSandwich, StableLearnedBloomFilter, PLBF, FastPLBF, SatoAtsu}. They achieved a better memory/accuracy trade-off than the original Bloom Filter.

Learned indexes are vulnerable to data update operations in general. This is because if the data distribution is distorted by updating operations, the accuracy of the machine learning model will decrease, and search performance will decrease, too. To address this problem, learned indexes that support efficient updating operations have been proposed~\cite{ALEX, FITing-Tree, PGM, LIPP}.

The above learned indexes can handle only one-dimensional data. On the other hand, learned indexes for multi-dimensional data have also been proposed. For example, Flood~\cite{Flood}, Tsunami~\cite{Tsunami}, Lisa~\cite{LISA}, RLR-tree~\cite{RLRTree}, Waffle~\cite{Waffle}, and so on~\cite{LearnedIndexSurvey}. Flood and Tsunami do not support data updating operations. Lisa, RLR-tree and Waffle support them, but there are no discussion of time complexity.

\section{Preliminary}\label{sec:Preliminary}
First, we define the problem setting. $N$ $D$-dimensional vectors ${\mathbf{v}_1, \dots, \mathbf{v}_N}$ are given, where $\mathbf{v}_n \in \mathbb{R}^D$. We denote the $d$-th dimension of $\mathbf{v}_n$ as $v_n[d] \in \mathbb{R}$. Initially, we construct a index based on the $N$ vectors. Then, we process $Q$ queries sequentially. Queries are provided in the following formats:

\begin{itemize}
    \item Search Query: $\mathbf{l} \in \mathbb{R}^D$ and $\mathbf{r} \in \mathbb{R}^D$ are provided. $\mathbf{l}$ and $\mathbf{r}$ represent the endpoints of the diagonal of the search range hyper-rectangle. We enumerate all vectors $\mathbf{v}_n$ contained within the data structure, such that for all $d \in \{1, 2, \dots, D\}$, $l[d] \leq v_n[d] \leq r[d]$ holds.
    \item Insert Query: A $D$-dimensional vector $\mathbf{v} \in \mathbb{R}^D$ is provided. We add it if this vector is not in the data structure. We do nothing if the same vector already exists within the data structure.
    \item Erase Query: A $D$-dimensional vector $\mathbf{v} \in \mathbb{R}^D$ is provided. We remove it if this vector exists within the data structure. We do nothing if the vector is not in the data structure.
\end{itemize}

Next, we explain the Flood algorithm~\cite{Flood}, which is the basis of the proposed method. Flood divides the $D$-dimensional space into approximately equal parts by $(D-1)$-dimensional grid cells. Given $D$-dimensional input vectors, Flood assigns them to the corresponding cell. Within each cell, Flood keeps vectors sorted using the value of the $D$-th dimension that is not used for the grid division. The parameters of Flood are the sort dimension and the number of cell partitions. Flood optimizes these parameters using a gradient descent method with a random forest regression model~\cite{RandomForest, RandomForestSKLearn}.

Flood uses sorted arrays to hold data in cells, but updating data on a sorted array is very expensive. We can make Flood updatable by replacing sorted arrays with B-trees. (Appendix~\ref{sec:SortedArrayvsBTree} discussed this overhead.) However, if the data distribution changes, the grid partitioning at the initialization becomes meaningless and the updatable Flood's search speed slows down greatly. Therefore, we propose an efficient data updating algorithm to solve this problem, using the updatable Flood as a baseline.

\section{Proposed Method: FlexFlood}\label{sec:FlexFlood}
\begin{figure}[tb]
  \begin{center}
   \includegraphics[keepaspectratio, width=\linewidth]{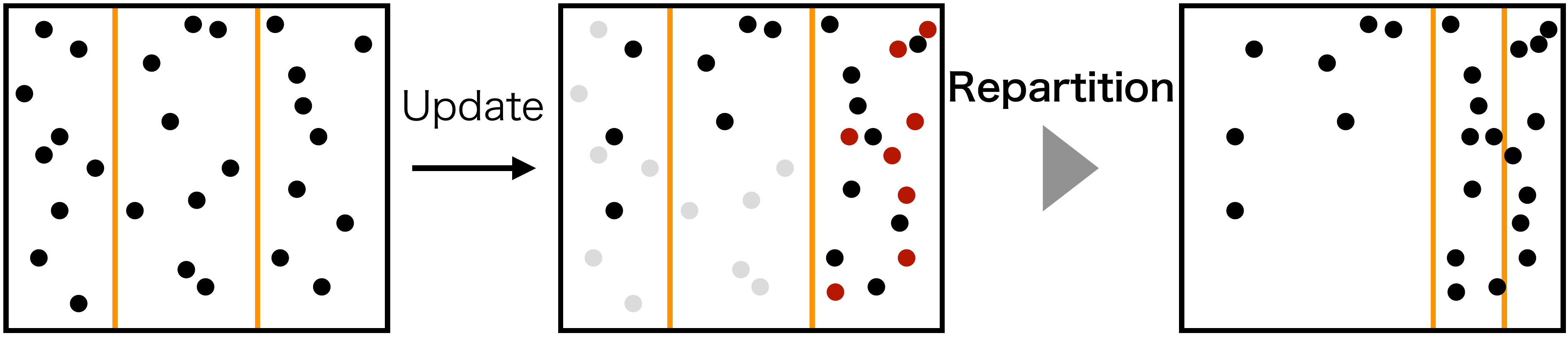}
  \end{center}
  \caption{Overview of our method ($D = 2$): Even if $N = 27$ data are equally divided into $x_1 = 3$ cells at the initialization, the distribution can become skewed due to data updating, and the updatable Flood slows down. We aim to ensure the search performance by partially re-partitioning the cells.}
  \label{DataSkewAndRepartition}
\end{figure}

Based on the updatable Flood, we propose FlexFlood, a data updating algorithm that ensures fast search even if data distribution changes. Figure~\ref{DataSkewAndRepartition} is an overview of FlexFlood. Remember Flood constructs cells so that for each axis $d$, the total number of data in the cell with the same $d$-dimensional value is $\frac{N}{x_d}$ (an equal number of data for each cell). Here, $x_d$ is the number of cell partitions on axis $d$. When the number of vectors in each cell deviates far from $\frac{N}{x_d}$ due to data updating, the search of the updatable Flood slows down greatly. Therefore, FlexFlood re-partitions around the cells as follows. (Appendix~\ref{sec:ThresholdforRepartitioning} discussed thresholds for re-partitioning in more detail.)

\begin{itemize}
    \item If the cell contains more than $\frac{2N}{x_d}$ vectors, we split the cell into two.
    \item If the cell contains less than $\frac{N}{3x_d}$ vectors and the neighboring cell contains less than $\frac{7N}{6x_d}$ vectors, we merge the two cells.
    \item If the cell contains less than $\frac{N}{3x_d}$ vectors and the neighboring cell contains more than $\frac{7N}{6x_d}$ vectors, we equalize the number of vectors in the two cells.
\end{itemize}

Re-partitioning in this way keeps the number of vectors in the cell constant and keeps the search speed fast. However, since the above re-partition requires a large amount of movement of data within the cells, it is not obvious whether the update operation can be performed efficiently.

We therefore analyzed the time complexity of this algorithm. Based on the insight that re-partitions with very high computational costs occur only with sufficiently low frequency, we found that the computational complexity of the updating operation is not large when considered in terms of amortization. Under the two assumptions that (1) the data increases at a pace that can be regarded as constant and (2) $\prod_{d=1}^{D} x_d \sum_{d=1}^{D} x_d \leq D N \log N$, we proved that the updating operation of FlexFlood is amortized at $O(D \log N)$. (See Appendix~\ref{sec:AmortizedTimeComplexityAnalysis} for detailed analysis.)

The update operation of the updatable Flood is $O(\log N)$, which is $D$ times faster than ours. Therefore, we can interpret FlexFlood as an algorithm that ensures high search performance even when the data distribution is skewed, instead of sacrificing up to $D$ times the complexity of the update operations.

\section{Experiment}\label{sec:Experiment}
We evaluate FlexFlood. The runtime environment is Intel Core i7-11800H, 8 cores, 2.3 GHz, 32 GB memory. The four data structures used for comparison and their implementations are shown below.

\begin{itemize}
    \item Self-Balancing Kd-tree (SB-Kdtree)~\cite{KdTree, ScapegoatTrees}: We implemented it in C++.
    \item R-tree~\cite{RTree}: We used C++ boost::geometry::index::rtree~\cite{BoostRTree}.
    \item Updatable Flood: We implemented Flood in C++, then we replaced Flood's sorted arrays with B-trees published by Google~\cite{GoogleBTree}.
    \item FlexFlood: We added cell re-partitioning algorithm to the updatable Flood.
\end{itemize}

As a dataset, we used (1) normal distribution, (2) Stock Price dataset~\cite{StockPrice}, and (3) Open Street Map dataset~\cite{OpenStreetMap}. See Appendix~\ref{sec:DatasetDetails} for details on data and query generation methods.

\section{Result}\label{sec:Reslut}
\begin{figure*}
    \centering
    \begin{subfigure}[tb]{0.32\textwidth}
        \centering
        \includegraphics[width=\textwidth]{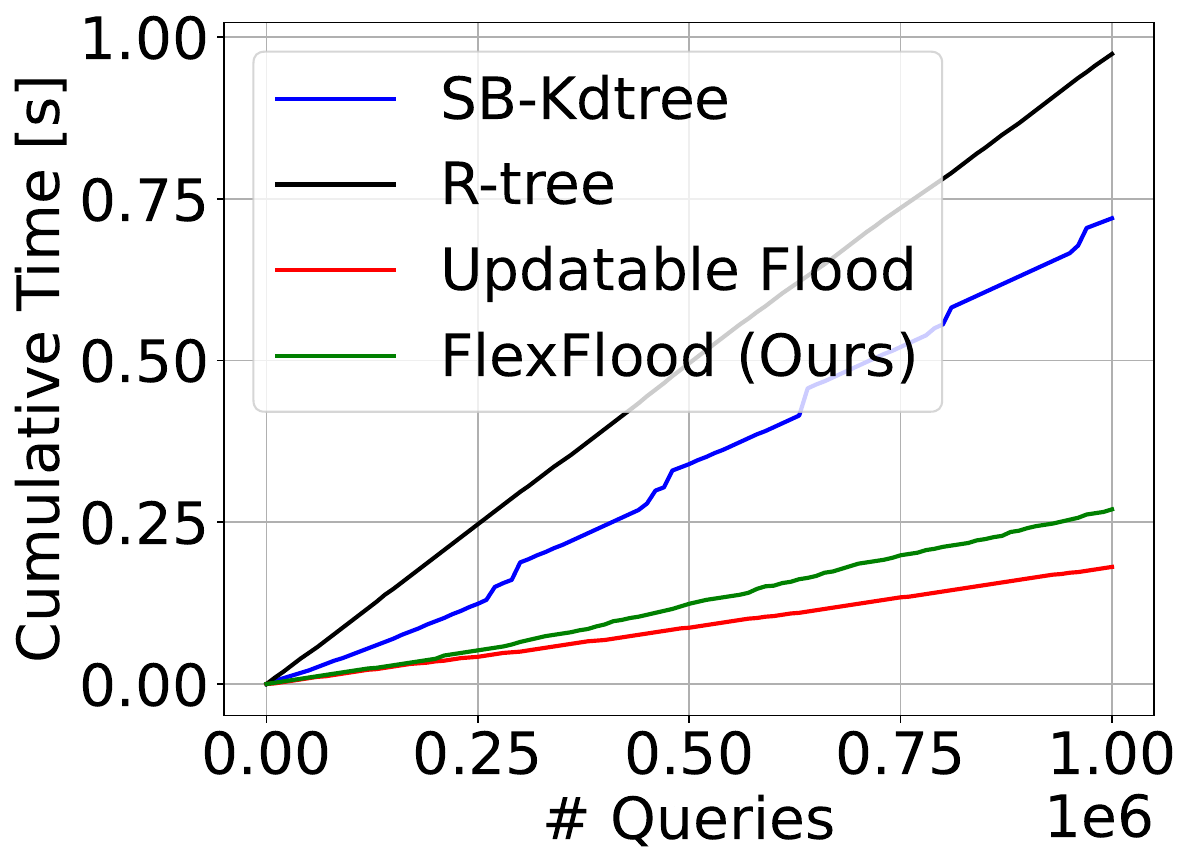}
        \caption{Normal Distribution Update}
        \label{Normal3DUpdate}
    \end{subfigure}
    \hfill
    \begin{subfigure}[tb]{0.32\textwidth}
        \centering
        \includegraphics[width=\textwidth]{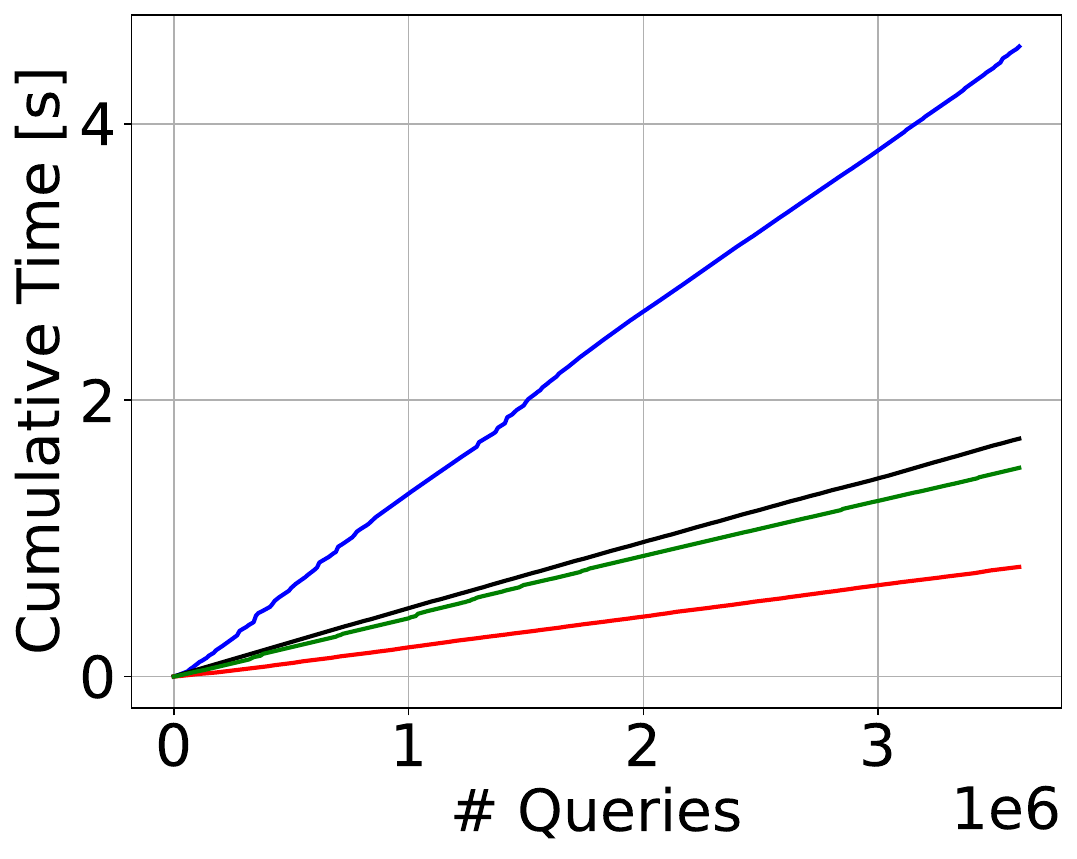}
        \caption{Stock Price Update}
        \label{Stock4DUpdate}
    \end{subfigure}
    \hfill
    \begin{subfigure}[tb]{0.32\textwidth}
        \centering
        \includegraphics[width=\textwidth]{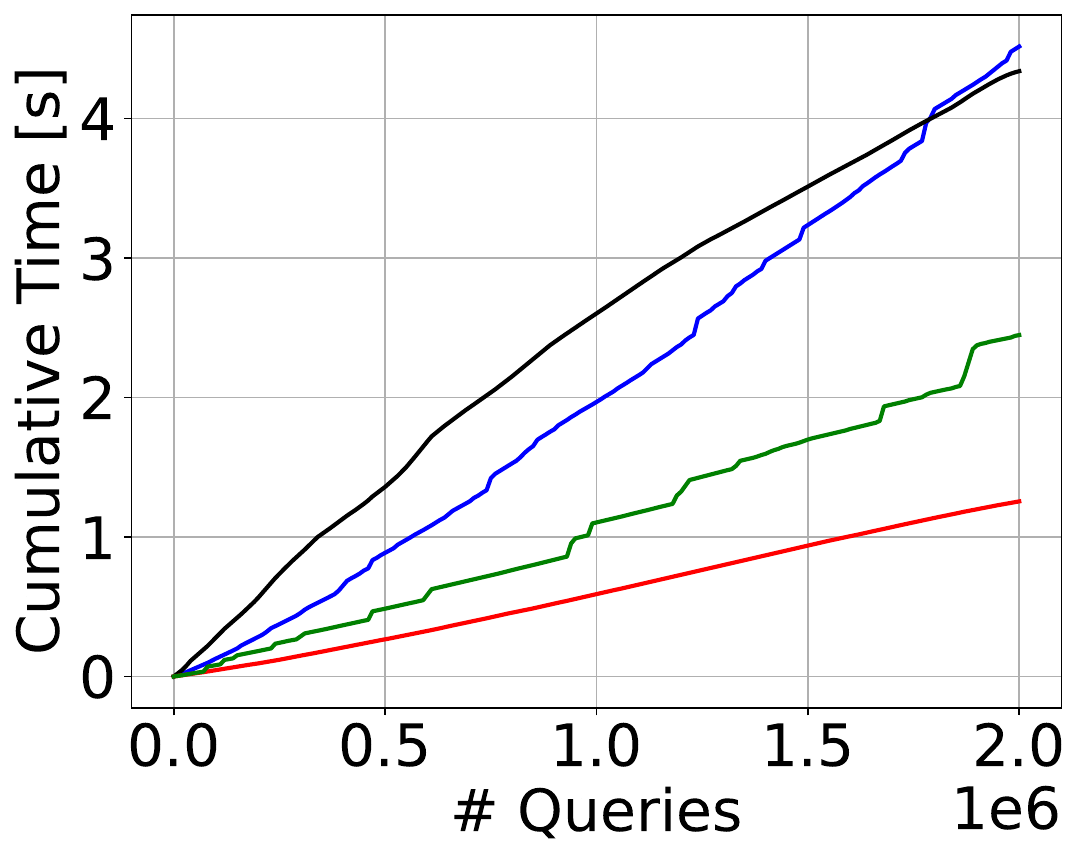}
        \caption{Open Street Map Update}
        \label{OSM5DUpdate}
    \end{subfigure}
    \\
    \begin{subfigure}[tb]{0.32\textwidth}
        \centering
        \includegraphics[width=\textwidth]{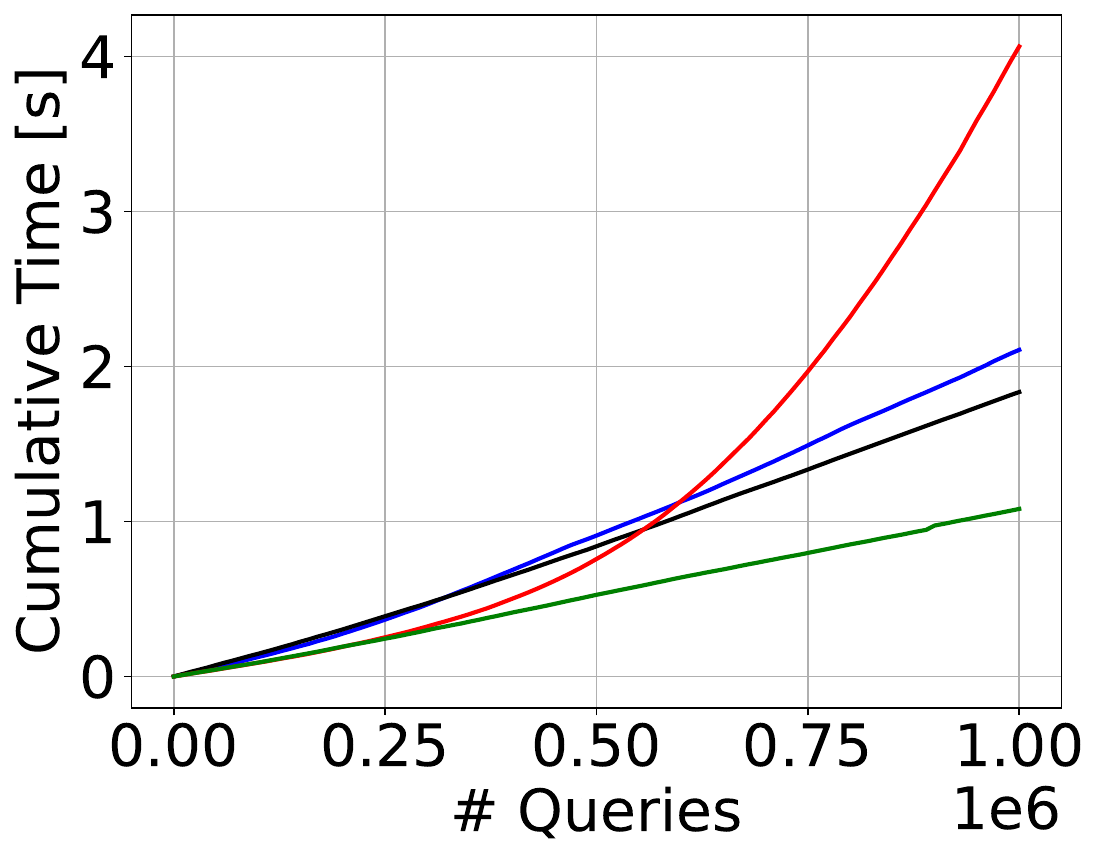}
        \caption{Normal Distribution Search}
        \label{Normal3DRead}
    \end{subfigure}
    \hfill
    \begin{subfigure}[tb]{0.32\textwidth}
        \centering
        \includegraphics[width=\textwidth]{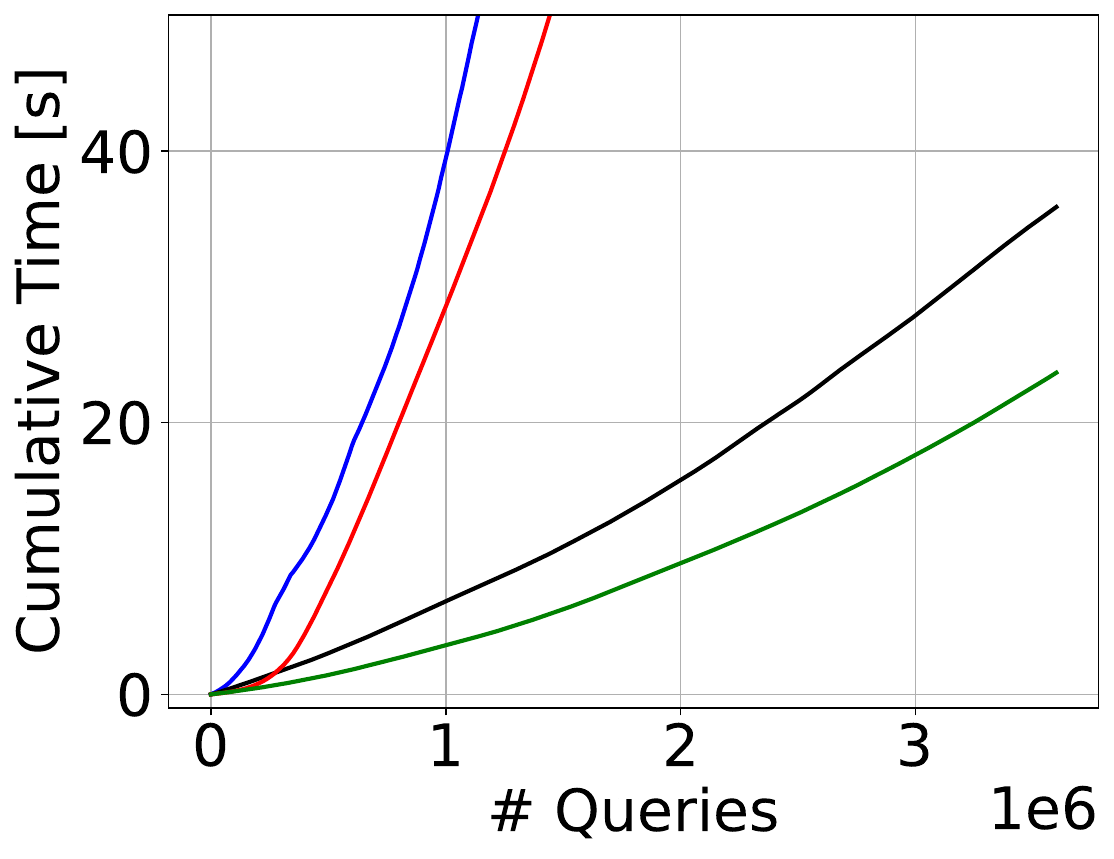}
        \caption{Stock Price Search}
        \label{Stock4DRead}
    \end{subfigure}
    \hfill
    \begin{subfigure}[tb]{0.32\textwidth}
        \centering
        \includegraphics[width=\textwidth]{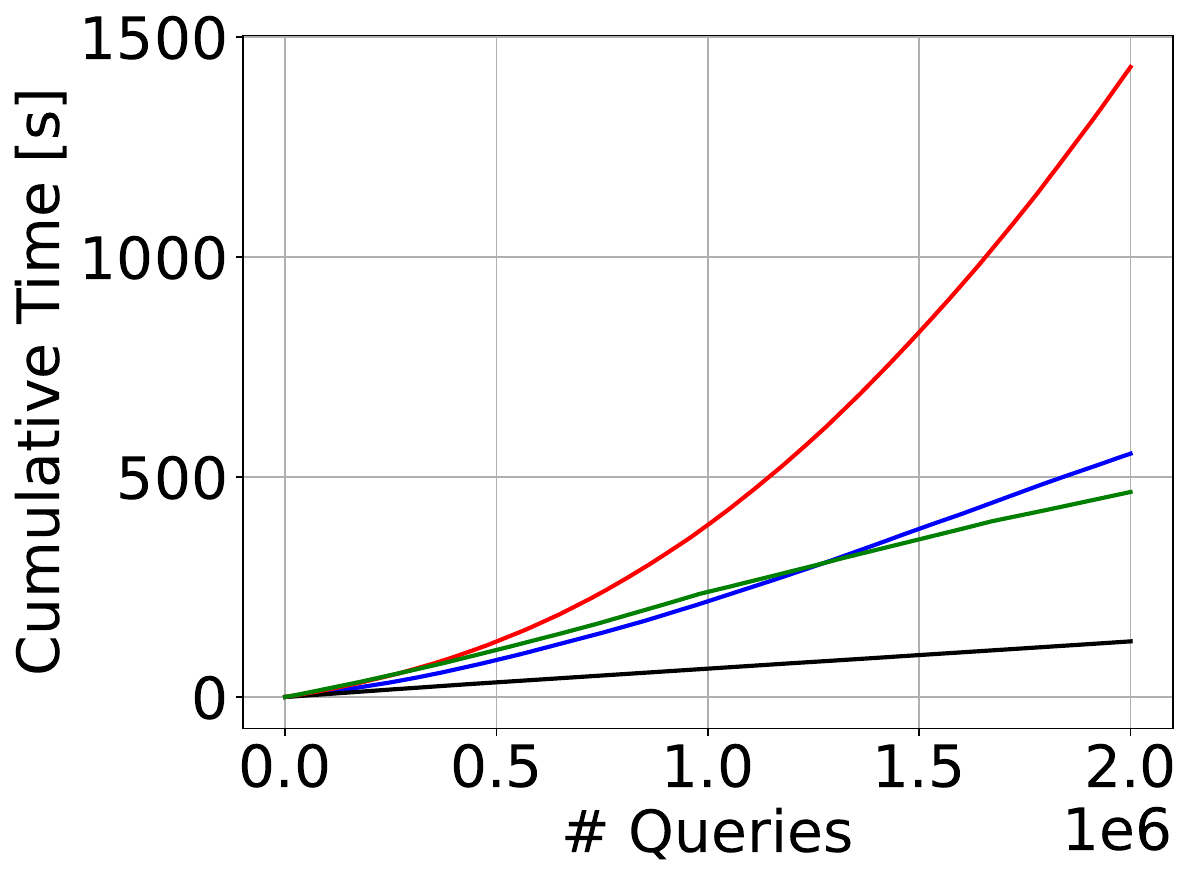}
        \caption{Open Street Map Search}
        \label{OSM5DRead}
    \end{subfigure}
\caption{Experimental results: The upper panel shows the update queries, and the lower panel shows the results for the search queries. (Lower is better.)} 
\label{fig:Result}
\end{figure*}

Figures~\ref{fig:Result} illustrate the cumulative query processing time for each data structure for each dataset. The number of cell partitions and whether the condition $\prod_{d=1}^{D} x_d \sum_{d=1}^{D} x_d \leq D N \log N$ was satisfied or not in each dataset are shown in Table~\ref{PartitionNumber}.

First, looking at Table~\ref{PartitionNumber}, we can see that the assumption (2) was always satisfied within the range of the experiments conducted here. Therefore, we can assume that the amortized time complexity of the data updating operation of FlexFlood is $O(D \log N)$ for practical purposes.

We then turn focus on Figures~\ref{Normal3DUpdate}, ~\ref{Stock4DUpdate}, ~\ref{OSM5DUpdate}, the results of the update queries. Comparing FlexFlood with the classical data structures such as SB-Kdtree and R-tree, FlexFlood processes update queries about 1.1 to 2.9 times faster. FlexFlood is better than the SB-Kdtree and R-tree for all datasets. Comparing FlexFlood with the updatable Flood, FlexFlood requires at most 2.0 times longer runtime. Remember that in theory, the cell re-partitioning algorithm takes about $D$ times longer runtime than the updatable Flood's data updating operation in the worst case. In light of this, we believe that FlexFlood is not only theoretically guaranteed to be computationally feasible but also fast enough for practical use (e.g., FlexFlood could be $D=5$ times slower than the updatable Flood for 5D Open Street Map dataset, but Figure~\ref{OSM5DUpdate} shows that FlexFlood is only about 2.0 times slower in practice).

Finally, we refer to Figures~\ref{Normal3DRead}, ~\ref{Stock4DRead}, ~\ref{OSM5DRead}, the results of the search queries. Comparing FlexFlood with the updatable Flood, FlexFlood processes search queries about 3.3 to 10 times faster, outperforming the updatable Flood on all datasets. Comparing FlexFlood with SB-Kdtree and R-tree, FlexFlood processes queries on the normal distribution dataset and the Stock Price dataset about 1.2 to 12 times faster. On the Open Street Map dataset, however, FlexFlood is slower than the R-tree although FlexFlood is faster than SB-Kdtree. Regarding this result, the benchmark paper~\cite{BenchmarkingLearnedIndexes} pointed out that the learned indexes perform poorly compared to classical data structures on the Open Street Map dataset because it lacks local structure, making them difficult to learn. Figure~\ref{OSM5DRead} also shows that the slope of the curve near the origin for the updatable Flood is steeper than that of SB-Kdtree and R-tree. In light of this fact, we interpret this result as consistent with the results of previous studies.

\begin{table}[tb]
    \centering
    \begin{tabular}{@{}llllc@{}} \toprule
        Dataset & $\{x_d\}_{d=1}^D$ & $\prod_{d=1}^{D} x_d \sum_{d=1}^{D} x_d$ & $D N \log N$ & Assumption (2) \\ \midrule
        3D Normal Distribution & $\{21, 17, 1\}$ & $1.4 \cdot 10^4$ & $5.0 \cdot 10^6$ & \checkmark \\
        4D Stock Price & $\{19, 19, 1, 39\}$ & $1.1 \cdot 10^6$ & $1.4 \cdot 10^7$ & \checkmark \\
        5D Open Street Map & $\{17, 17, 13, 23, 1\}$ & $6.1 \cdot 10^6$ & $1.0 \cdot 10^8$ & \checkmark \\ \bottomrule
    \end{tabular}
    \caption{Number of cell partitions and whether the conditional expression was satisfied or not}
    \label{PartitionNumber}
\end{table}

\section{Conclusion}\label{sec:Conclusion}
By adaptively reconstructing the internal structure of Flood, we proposed FlexFlood, which supports efficient data updating. Experimental results show that FlexFlood does not reduce the search speed and has advantages over classical data structures. Furthermore, we proved that the amortized time complexity of data updating is $O(D \log N)$ under two experimentally valid assumptions.

On the other hand, FlexFlood loses the optimality guarantees regarding the sort dimension and the number of cell divisions after the data update. Therefore, it may be possible to ensure even faster search by periodically relearning the distribution. (Appendix~\ref{sec:ReInitialization} discusses this in more detail.)

\bibliographystyle{plainnat}
\newpage
\bibliography{references}
\newpage

\appendix

\section{Sorted Array vs B-tree}\label{sec:SortedArrayvsBTree}
The updatable Flood replaces the Flood's sorted arrays with B-trees, which slows down the search. Therefore, we conducted comparative experiments on workloads where no update queries existed. The data structures used for the comparison are (1) Self-Balancing Kd-tree (SB-Kdtree), (2) R-tree, (3) Flood, and (4) updatable Flood. Note that FlexFlood is exactly the same as the updatable Flood if there are no update queries. The dataset used for the experiments are (1) normal distribution, (2) Stock Price dataset, and (3) Open Street Map dataset. We initialized each data structure, and performed $10^4$ search queries to measure the processing time per search query.

\begin{figure*}
    \centering
    \begin{subfigure}[tb]{0.325\textwidth}
        \centering
        \includegraphics[width=\textwidth]{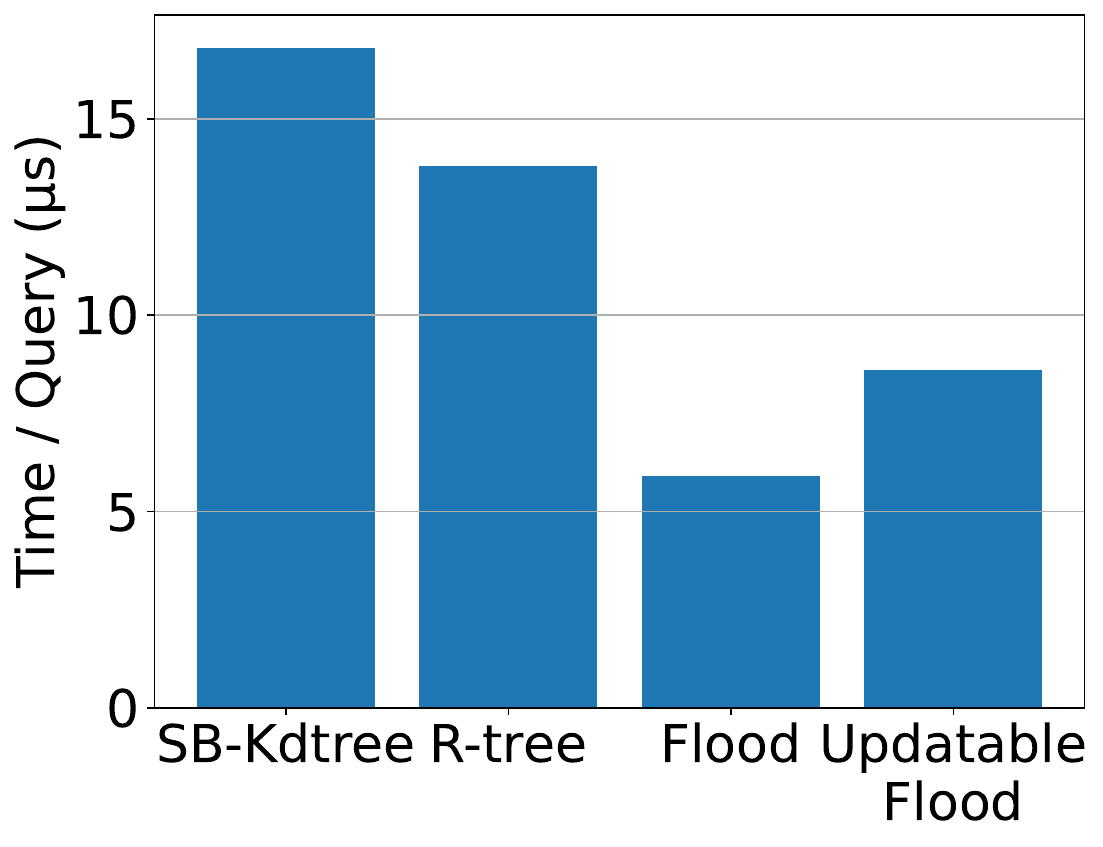}
        \caption{Normal Distribution}
        \label{Normal3DReadOnly}
    \end{subfigure}
    \hfill
    \begin{subfigure}[tb]{0.325\textwidth}
        \centering
        \includegraphics[width=\textwidth]{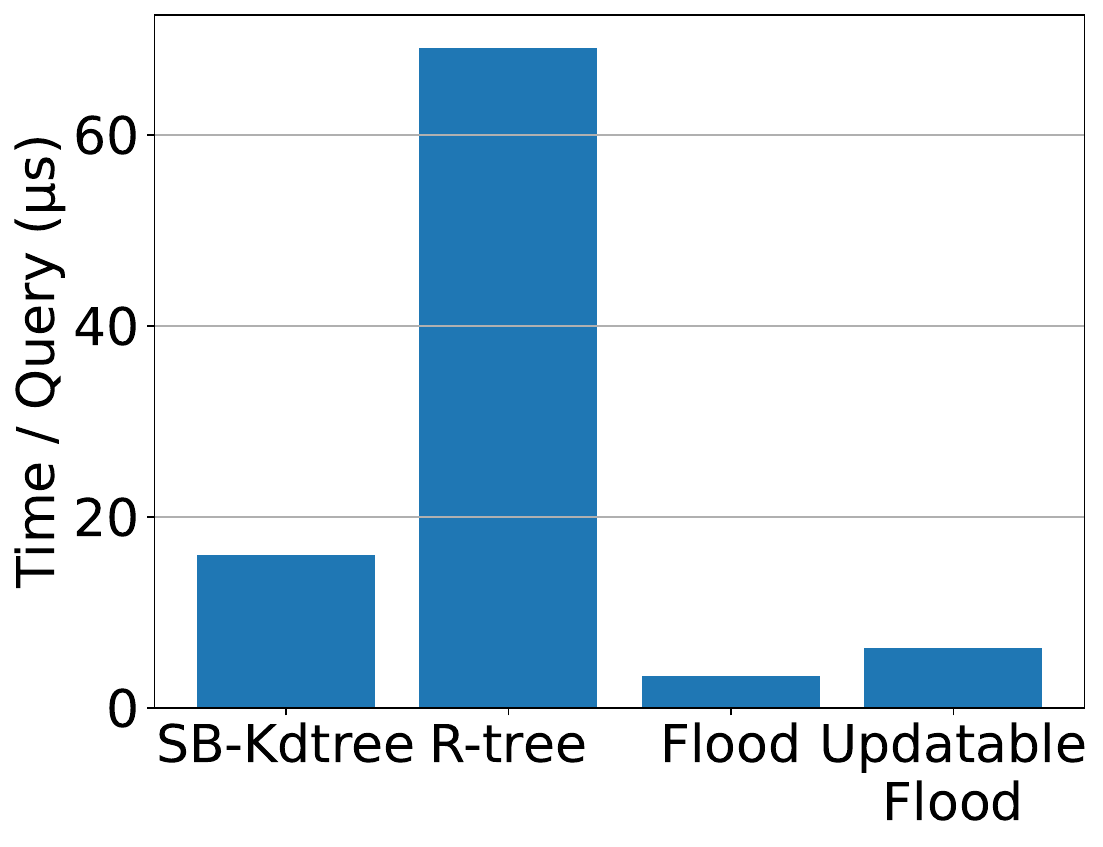}
        \caption{Stock Price}
        \label{Stock4DReadOnly}
    \end{subfigure}
    \hfill
    \begin{subfigure}[tb]{0.325\textwidth}
        \centering
        \includegraphics[width=\textwidth]{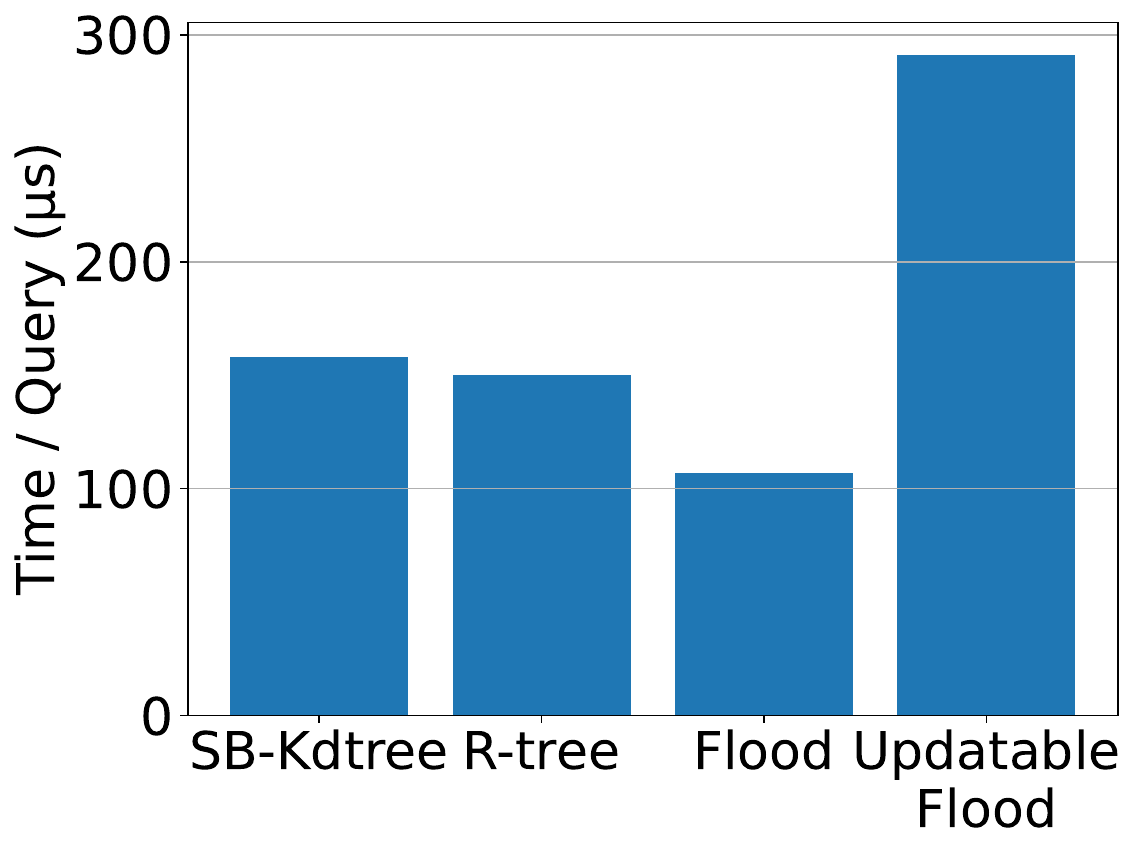}
        \caption{Open Street Map}
        \label{OSM5DReadOnly}
    \end{subfigure}
\caption{Processing time per search query.} 
\label{fig:ReadOnlyResult}
\end{figure*}

The experimental results are in Figure~\ref{fig:ReadOnlyResult}. The search speed of the updatable Flood is about 1.5 to 2.7 times slower than that of Flood. Therefore, we should use regular Flood for workloads where it is known in advance that there will be no update queries at all. However, for many datasets, updatable Flood achieves faster search than classical data structures. Therefore, we believe that updatable Flood is worth using for workloads where update queries are likely to be present.

\section{Threshold for Re-partitioning}\label{sec:ThresholdforRepartitioning}
The thresholds for the re-partitioning algorithm introduced in Section~\ref{sec:FlexFlood} are hyperparameters. In this section, we discuss the intuitive reasons for determining the thresholds as in Section~\ref{sec:FlexFlood} and how the performance changes when the thresholds are varied.

\subsection{Threshold Selection Criteria}
We explain why the threshold for ``split'' in Section~\ref{sec:FlexFlood} is set to $\frac{2N}{x_d}$ and the threshold for ``merge'' and ``equalize'' to $\frac{N}{3x_d}$. The purpose of cell re-partitioning is to maintain the number of data in cells at a baseline value $\frac{N}{x_d}$ when the number of data in a particular cell increases or decreases too much. Since ``split'' halves the number of data in a cell, it is efficient to perform a ``split'' when the number of data in a cell is $\frac{2N}{x_d}$. For the same reason, it seems intuitive that ``merge'' and ``equalize'' should be performed when the number of data in a cell reaches $\frac{N}{2x_d}$. However, the number of data in the neighboring cell is greater than $\frac{N}{2x_d}$ and less than $\frac{2N}{x_d}$. Therefore, the number of data in the cell after ``merge'' and ``equalize'' is considered to be more than $\frac{N}{x_d}$. For these reasons, we selected $\frac{N}{3x_d}$ as the threshold for ``merge'' and ``equalize'', which is slightly smaller than $\frac{N}{2x_d}$. This is expected to bring the number of data in the cell after ``merge'' and ``equalize'' closer to $\frac{N}{x_d}$.

Also, we explain why the threshold for switching between ``merge'' and ``equalize'' is set to $\frac{7N}{6x_d}$. This is because $\frac{7N}{6x_d} = \frac{\frac{N}{3x_d} + \frac{2N}{x_d}}{2}$. When ``merge'' or ``equalize'' is performed, the number of data in the neighboring cell is between $\frac{N}{3x_d}$ and $\frac{2N}{x_d}$. Therefore, we switch between ``merge'' and ``equalize'' at the intermediate value.

\subsection{Performance Variation with Thresholds}
In order to confirm the appropriateness of the thresholds in Section~\ref{sec:FlexFlood}, we describe the results of the experiments in which FlexFlood's performance changed when the thresholds were changed. We conducted the experiments by varying the coefficients of $\frac{N}{x_d}$ on the threshold of ``split'' in the range $\{1.5, 1.6, 1.7, 1.8, 1.9, 2.0, 2.1, 2.2, 2.3, 2.4, 2.5\}$ and on the threshold of ``merge'' and ``equalize'' in the range $\{0.1, 0.15, 0.2, 0.25, 0.3, 0.35, 0.4, 0.45, 0.5, 0.55, 0.6\}$. The threshold for switching between ``merge'' and ``equalize'' was set to a value halfway between the two thresholds for re-partitioning. The experimental settings are the same as in Section~\ref{sec:Experiment}.

\begin{figure*}
    \centering
    \begin{subfigure}[tb]{0.48\textwidth}
        \centering
        \includegraphics[width=\textwidth]{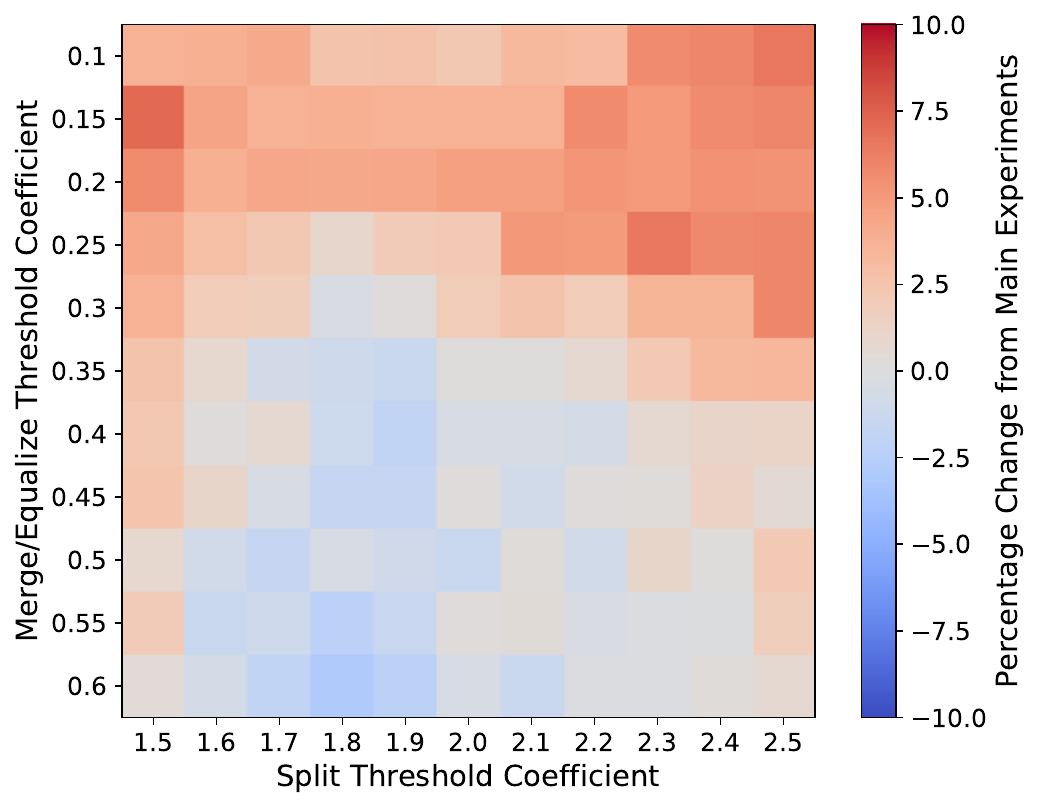}
        \caption{Normal Distribution Search}
        \label{NormalReadPartitioning}
    \end{subfigure}
    \hfill
    \begin{subfigure}[tb]{0.48\textwidth}
        \centering
        \includegraphics[width=\textwidth]{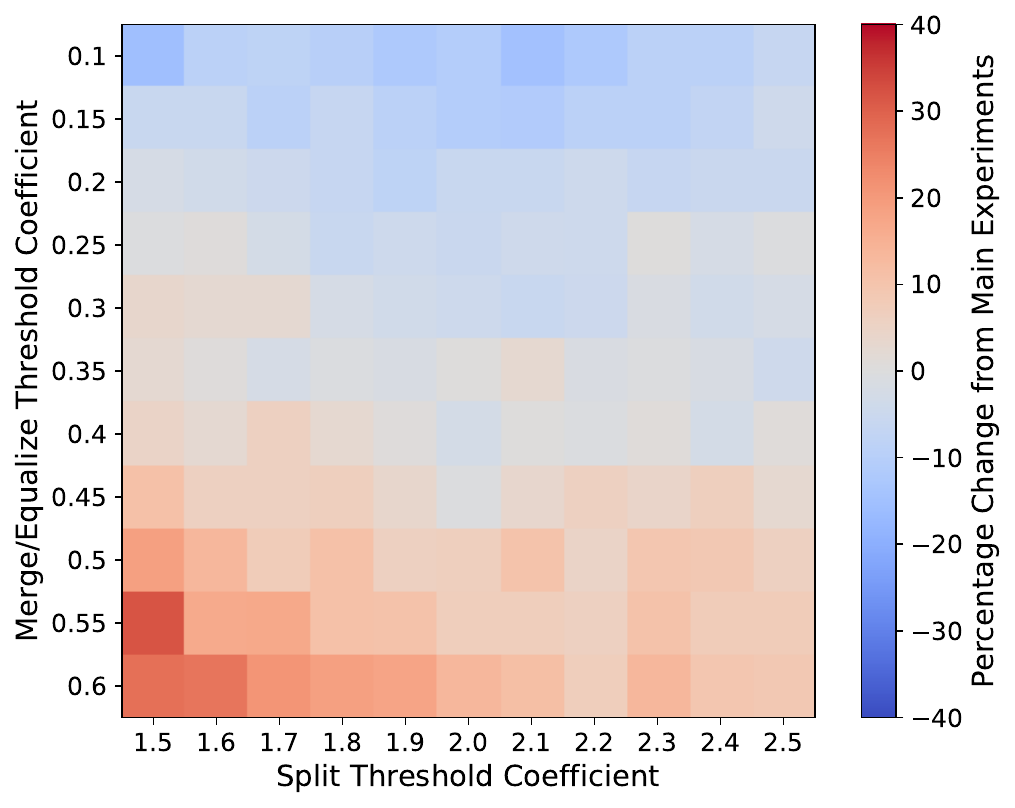}
        \caption{Normal Distribution Update}
        \label{NormalUpdatePartitioning}
    \end{subfigure}
    \\
    \begin{subfigure}[tb]{0.48\textwidth}
        \centering
        \includegraphics[width=\textwidth]{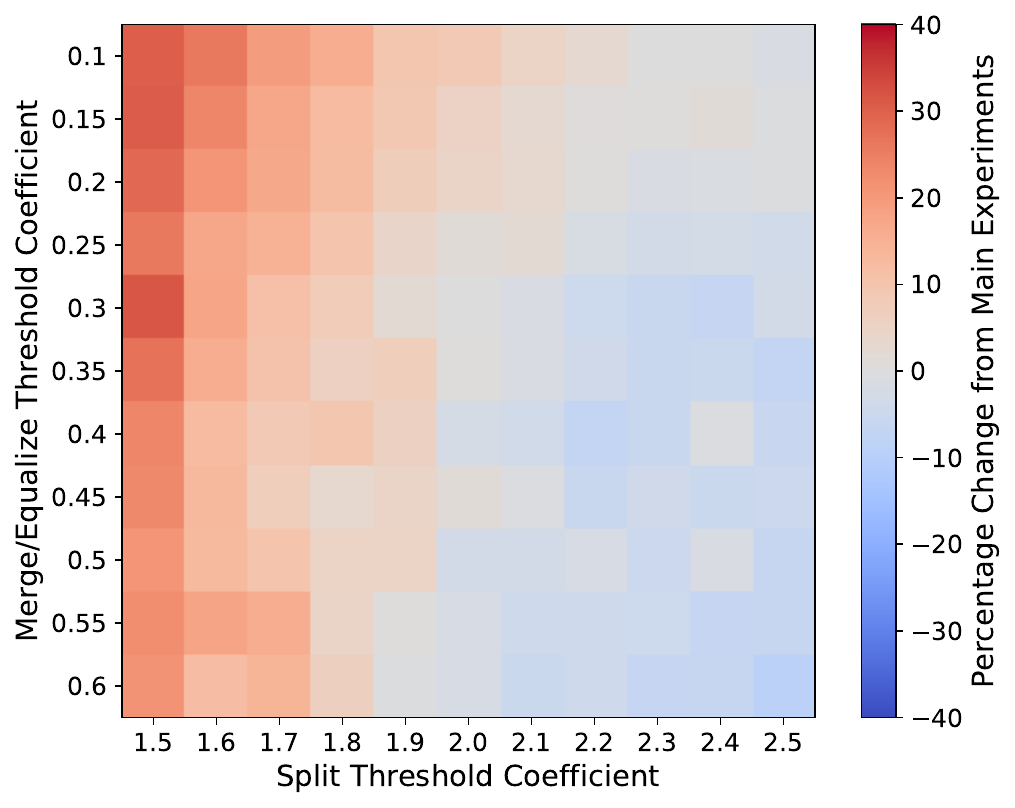}
        \caption{Stock Price Search}
        \label{StockReadPartitioning}
    \end{subfigure}
    \hfill
    \begin{subfigure}[tb]{0.48\textwidth}
        \centering
        \includegraphics[width=\textwidth]{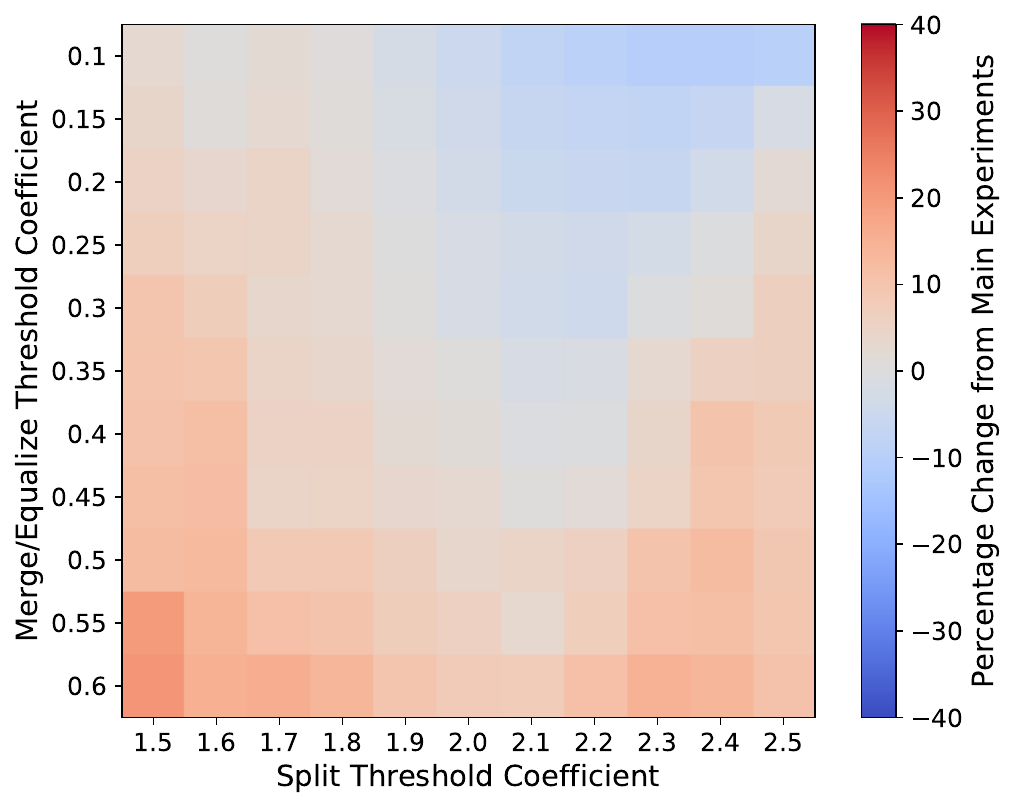}
        \caption{Stock Price Update}
        \label{StockUpdatePartitioning}
    \end{subfigure}
    \\
    \begin{subfigure}[tb]{0.48\textwidth}
        \centering
        \includegraphics[width=\textwidth]{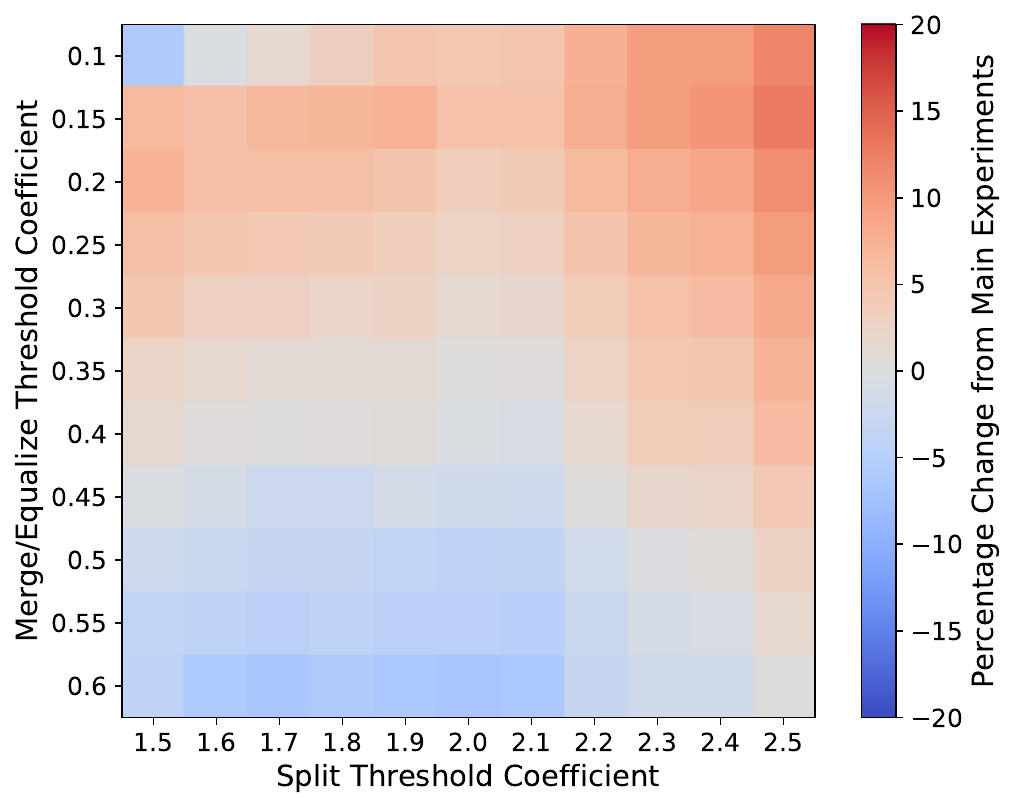}
        \caption{Open Street Map Search}
        \label{OSMReadPartitioning}
    \end{subfigure}
    \hfill
    \begin{subfigure}[tb]{0.48\textwidth}
        \centering
        \includegraphics[width=\textwidth]{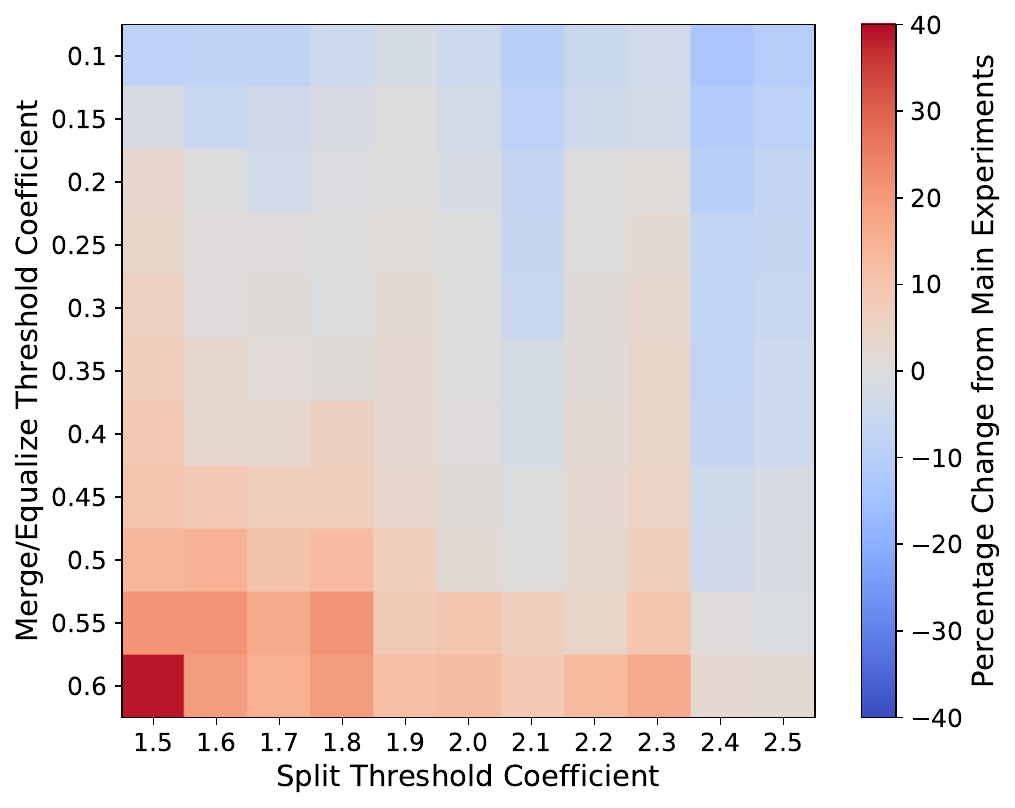}
        \caption{Open Street Map Update}
        \label{OSMUpdatePartitioning}
    \end{subfigure}
\caption{Heatmaps show percentage changes compared to experimental results in Section~\ref{sec:Reslut}. Blue squares mean better performance.} 
\label{fig:ResultPartitioningTest}
\end{figure*}

Figures~\ref{fig:ResultPartitioningTest} shows the results of the experiments in which search time and update time were measured separately. Red squares in the heatmap indicate a long search/update time, while blue squares indicate a short time. That is, blue squares mean better performance. In more detail, we calculated the percentage change compared to the search/update time when the threshold was set as section~\ref{sec:FlexFlood}. Figures~\ref{fig:ResultPartitioningTest} shows that there are few thresholds that are blue squares in the heatmaps for both search and update. Therefore, we believe that the thresholds of Section~\ref{sec:FlexFlood} are one of the best practical values of hyperparameters.

\section{Amortized Time Complexity Analysis}\label{sec:AmortizedTimeComplexityAnalysis}
We prove that the amortized time complexity of FlexFlood's update operation is $O(D \log N)$ under two assumptions. Let $X$ denote the total number of cells, that is, $\prod_{d=1}^{D} x_{d} = X$.

\subsection{Worst Time Complexity of Vector Insertion or Erasion}
First, we show that the worst time complexity of vector insertion or erasion is $O(\log N)$. Insertion or erasion of a vector $\mathbf{v} \in \mathbb{R}^D$ consists of two steps: (1) identifying the cell that should contain $\mathbf{v}$ and (2) inserting or erasing $\mathbf{v}$ for a B-tree within the cell. For (1), for each dimension $d \in \{1, 2, \dots, D\}$, we can perform a binary search on the set of cell boundary coordinates to determine where $v[d]$ should be placed. This takes $O(\sum_{d=1}^{D} \log x_{d}) = O(\log \prod_{d=1}^{D} x_{d}) = O(\log X)$. We can do (2) with worst $O(\log N)$ because we only insert or erase $\mathbf{v}$ into the B-tree that holds the data in the identified cell. The total computational cost of inserting or erasing vectors is $O(\log X + \log N)$. Here, $X < N$ is considered to be valid unless the learning of the distribution is very unsuccessful. This is because $X \geq N$ means that the number of cells is larger than the total number of data, which is obviously wasteful. Table~\ref{PartitionNumber} also shows that this is valid. Therefore, the worst time complexity of vector insertion or erasion is $O(\log N)$.

\begin{figure}[tb]
  \begin{center}
   \includegraphics[keepaspectratio, width=\linewidth]{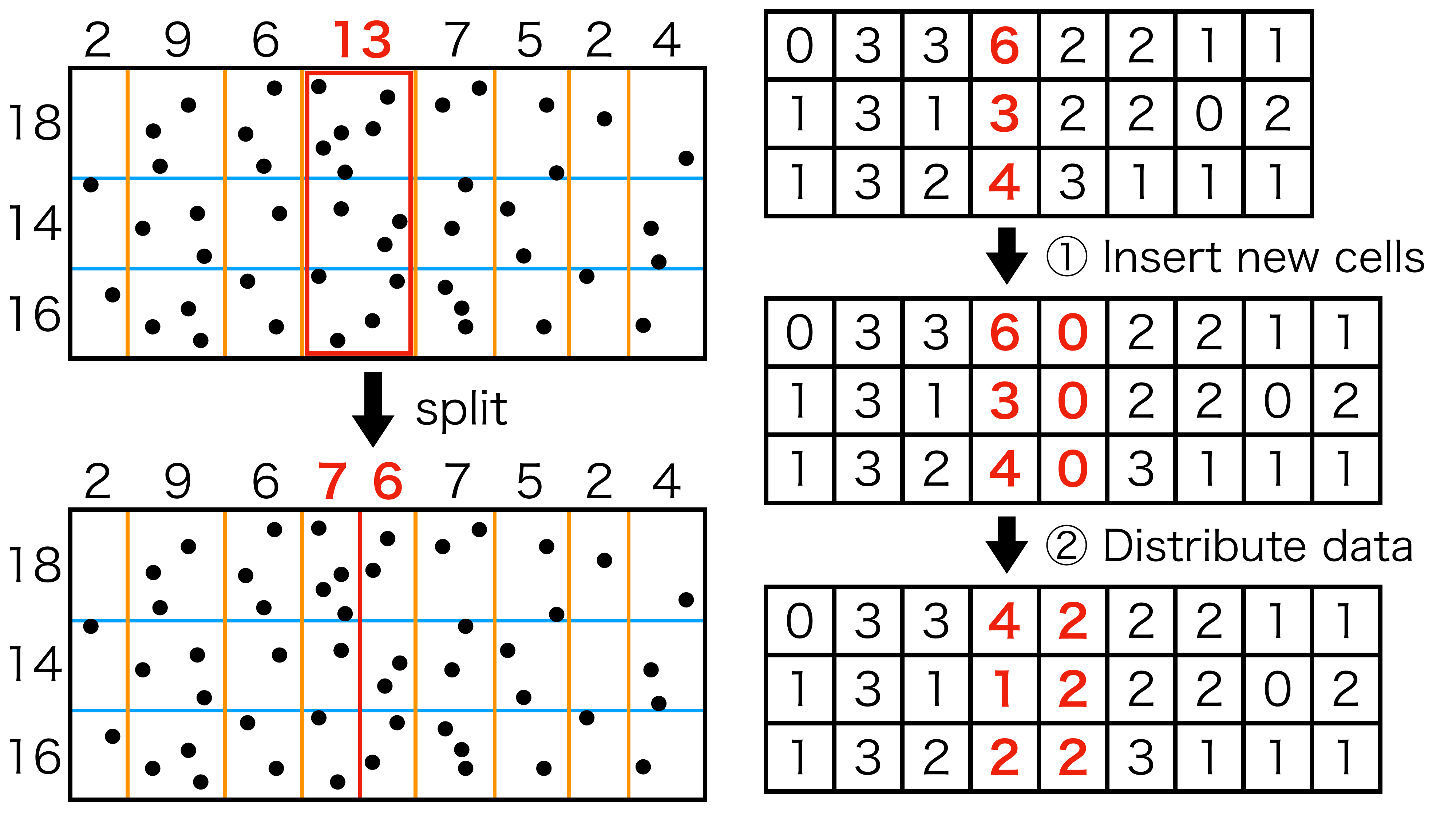}
  \end{center}
  \caption{Overview of ``split'' ($D = 3, x_1 = 8, x_2 = 3, X = 24, N = 48$): The cells outlined in red are subject to ``split'' because they contain 13 data in total. This value is larger than the ``split'' condition: $2 \cdot \frac{N}{x_1} = 2 \cdot \frac{48}{8} = 12$. We insert $\frac{X}{x_1} = \frac{24}{8} = 3$ new cells (B-trees), and distribute the 12 data points evenly between the old and new cells, with 6 points in each.}
  \label{SplitOverview}
\end{figure}

\subsection{Worst Time Complexity of Cell Re-partition}
Next, we discuss the worst time complexity of cell re-partition. Since cell re-partition occurs independently for each axis $d \in \{1, 2, \dots, D\}$, we consider each axis $d$ independently and sum up later. We assume the following two conditions.

\begin{itemize}
    \item The number of data $N$ increases at an approximately constant rate.
    \item $\prod_{d=1}^{D} x_d \sum_{d=1}^{D} x_d \leq D N \log N$
\end{itemize}

We define the first assumption as increasing the number of data by $\Delta \in [0, 1]$ per updating query. From a micro perspective, of course, each update operation either increases or decreases the data count by one. However, from a macro perspective, we assume that after performing $k$ update operations, the number of data $N$ can be approximated as $N = N_0 + k \Delta$ using the initial number of data $N_0$.

We discuss the worst-case time complexity of the ``split'' operation. An overview of the ``split'' is shown in Figure~\ref{SplitOverview}. ``Split'' first inserts empty cells (B-trees) in the appropriate location, and then distributes the data equally between the old and new cells. We can implement the insertion of empty cells by simultaneously inserting new $\frac{X}{x_d}$ cells while sliding (at most) $X$ cells that already exist. Thus, the insertion of empty cells costs $O(\frac{X}{x_d} + X) = O(X)$. For the data distribution, the erase from old B-trees and insert into new B-trees are performed $\frac{2N}{x_d}$ times in total (Remember the definition of the ``split''. Since $\Delta \geq 0$ holds, the ``split'' occurs when the number of data of the target cells is exactly $\frac{2N}{x_d}$, and the number of B-tree operations cannot be more than $\frac{2N}{x_d}$). Therefore, the time complexity of operating B-trees for data distribution is $O(\frac{N}{x_d} \log N)$. Thus, the worst time complexity of the ``split'' is $O(X + \frac{N}{x_d} \log N)$. Considering the ``merge'' and the ``equalize'' in the same way, the worst time complexities are $O(X + \frac{N}{x_d} \log N), O(\frac{N}{x_d} \log N)$, respectively.

\subsection{Amortized Time Complexity of Cell Re-partition}
However, under the assumption that $N = N_0 + k\Delta$, we can show that a re-partition occurs at most once every $O(\frac{N}{x_d})$. When $N = N_0 + k\Delta$, we insert $\frac{(1 + \Delta)k}{2}$ times and erase $\frac{(1 - \Delta)k}{2}$ times. ``Splits'' due to intensive insertions occur only $\frac{(1 + \Delta)k}{2} \cdot \frac{x_d - 2 \Delta}{N}$ times at most. This is because the ``split'' occurs when the number of data in the cell reaches exactly $\frac{2N}{x_d}$ (because $\Delta \geq 0$ holds), and once the ``split'' is executed, the number of data in the cell is reset to $\frac{N}{x_d}$. So, we need $\frac{N}{x_d - 2 \Delta}$ consecutive intensive insertions to execute the ``split'' again (Check $\frac{N}{x_d} + \frac{N}{x_d - 2 \Delta} = 2 \cdot \frac{N + \frac{N}{x_d - 2 \Delta}\Delta}{x_d}$). In the same way, we can say that the ``merge'' and the ``equalize'' by intensive erasion occur only $\frac{(1 - \Delta)k}{2} \cdot \frac{3x_d + \Delta}{2N}$ times at most. Moreover, even for cells that are not erased at all, the number of data in the cell may reach the threshold of ``merge'' or ``equalize'' because $N$ increases. About this case, we can prove that the number of computations is maximized when the ``merge'' or ``equalize'' occurs immediately when the number of data reaches the threshold. Therefore, the ``merge'' and the ``equalize'' due to increasing data can occur only $x_d \cdot \frac{k}{\frac{2N}{\Delta}}$ times at most. From the above, the upper bound on the number of cell re-partitions that occur after $k$ update operations is $\frac{k((5 + \Delta)x_d - 3\Delta - 5\Delta^2)}{4N}$. That is, the cell re-partitioning occurs at most once every $\frac{4N}{(5+\Delta)x_d - 3\Delta - 5\Delta^2} = O(\frac{N}{x_d})$ times.

Remember the worst-case complexity of the cell re-partitioning is $O(X + \frac{N}{x_d} \log N)$. This happens once every $O(\frac{N}{x_d})$ times, so the amortized time complexity of the cell re-partitioning is $O(\frac{X}{N} x_d + \log N)$. Since we have considered each axis $d$ independently, the amortized time complexity of the overall cell re-partitioning is $O(\frac{X}{N} \sum_{d=1}^{D} x_d + D \log N)$ by summing them up. Here, from the second assumption $\prod_{d=1}^{D} x_d \sum_{d=1}^{D} x_d \leq D N \log N$, $\frac{X}{N} \sum_{d=1}^{D} x_d < D \log N$ holds. Therefore, it is proved that the amortized time complexity of the cell re-partitioning is $O(D \log N)$.

\subsection{Limitation}
We proved that the amortized time complexity of the FlexFlood's update operation is $O(D \log N)$ when $N \simeq N_0 + k \Delta$ and $\prod_{d=1}^{D} x_d \sum_{d=1}^{D} x_d \leq D N \log N$ are assumed to hold. It remains to be discussed how realistic these assumptions are. In the real world, data basically tends to increase. However, for example, there may be cases where periods of rapid data growth alternate with periods of slower growth on an annual cycle. It is very important to observe how FlexFlood performs in such cases. If the performance drops significantly, we should develop a hybrid approach that balances these periods. Furthermore, although $\prod_{d=1}^{D} x_d \sum_{d=1}^{D} x_d \leq D N \log N$ always held in our experiment, it may not hold for some datasets. We should observe how much the update speed decreases due to the breakdown of this assumption. Furthermore, it would be interesting to identify common characteristics among datasets where this assumption breaks down.

\section{Dataset Details}\label{sec:DatasetDetails}
We describe the details of the three dataset used in the experiments. In addition, we also introduce the method for generating queries.

\subsection{Normal Distribution Dataset}
Dataset 1 is a Normal Distribution Dataset. We tested $D = 3$ as the number of dimensions of the normal distribution. As the initial data, we independently generated $10^5$ data. Each data was generated independently for each axis according to a normal distribution with $\mu = 3 \cdot 10^8$, $\sigma = 10^8$. Note that $\mu$ is the mean and $\sigma$ is the standard deviation of the normal distribution.

The number of queries we generated is $2 \cdot 10^6$, with an update and search queries alternating every $10^4$ queries. The $i$-th query is an update query if $\lfloor \frac{i}{10^4} \rfloor \equiv 0 \pmod 2$, and a search query otherwise. (Note that $\lfloor x \rfloor$ denotes the largest integer less than or equal to $x$.)

When we generated search queries, we first generated a hyper-rectangle, a search region. The length of one side of the hyper-rectangle was independently set to a random value less than $3 \cdot 10^8$. We placed that hyper-rectangle uniformly at random inside the hyper-rectangle whose diagonals are $(0, 0, 0)$ and $(10^9, 10^9, 10^9)$, and we used this as a search query.

The update query has a $50\%$ chance of being selected as an insert query and a $50\%$ chance of being selected as an erase query. We generated the insertion queries according to a normal distribution with $\mu = 3 \cdot 10^8 + 4 \cdot 10^8 \cdot \frac{i}{2 \cdot 10^6}$, $\sigma = 10^8$ independently for each axis when the query was the $i$-th from the first. This simulates the gradual change of the data distribution. We generate deletion queries by randomly selecting one of the data currently in the data structure.

\subsection{Stock Price Dataset}
Dataset 2 is the Stock Price Dataset~\cite{StockPrice} used by~\cite{Tsunami}. we tested $D = 4$ as the number of dimensions. Of the approximately $2 \cdot 10^7$ of data included in the dataset, we randomly chose $2 \cdot 10^6$ data and used them. Each data has four attributes: lowest price, highest price, volume, and date.

Out of $2 \cdot 10^6$ data points, we took $2 \cdot 10^5$ with the oldest dates as the initial data. We generated $7.2 \cdot 10^6$ queries, with the search and update query order matching Dataset 1. Hyper-rectangles were generated randomly for search queries, searching approximately 0.1$\%$ of the data per query. Insert queries were generated by selecting the oldest data not yet in the structure, and erase queries by selecting the oldest data in the structure. This simulated keeping the last $2 \cdot 10^5$ data points and evaluated each data structure’s robustness to real-world data distribution shifts over time.

\subsection{Open Street Map Dataset}
Dataset 3 is the Open Street Map Dataset~\cite{OpenStreetMap} used by~\cite{Flood}. We tested $D = 5$ as the number of dimensions. Of the approximately $10^8$ of data, we randomly extracted $2 \cdot 10^6$ data and used them. Each data set has five attributes: ID, version, date, latitude, and longitude.

First, we clustered the dataset into two groups of $10^6$ each using K-means, and used one cluster as the initial data. We generated $2 \cdot 10^6$ queries, with the search and update query order matching Dataset 1. We generated hyper-rectangles randomly for search queries so that approximately 0.3$\%$ of the data were searched per query. Insert queries were generated by randomly selecting data from the cluster not selected as the initial data, while erase queries were generated by randomly selecting data from the initial cluster.

\section{Re-initialization}\label{sec:ReInitialization}
There is a delta-buffer method according to the survey paper on learned multi-dimensional indexes~\cite{LearnedIndexSurvey}. It stores data updating in a small array and periodically merges them with the data structure. We compare the performance of FlexFlood with that of a delta-buffered version of Flood.

In designing the delta-buffer Flood, it is computationally too expensive to incorporate re-learning of parameters (sort dimension and the number of cell partitions). In our implementation, learning the distribution took 32 seconds for the normal distribution dataset, 46 seconds for the Stock Price dataset, and 138 seconds for the Open Street Map dataset. Since there are about $10^6$ update queries in our workload, we need to re-learn more than $10^2$ times even if we re-learn every $10^4$ times. The total re-learning time would then be about $10^4$ seconds $\simeq$ 3 hours, which is too long as shown in Figures~\ref{fig:Result}. Therefore, we adopt the method of re-initializing the data structure without changing the parameters.

In this experiment, we used Flood10000, which re-initializes the entire structure after every $10^4$ data updates, and used Flood20000 with re-initialization after every $2 \cdot 10^4$ updates, in addition to the four data structures described in Section~\ref{sec:Experiment}. As noted in Appendix~\ref{sec:DatasetDetails}, search queries and update queries come alternately, $10^4$ each. Thus, Flood10000 is re-initialized for each set of update queries, and Flood20000 stores the first set of update queries in the buffer and re-initializes after the second set.

\begin{figure*}
    \centering
    \begin{subfigure}[tb]{0.32\textwidth}
        \centering
        \includegraphics[width=\textwidth]{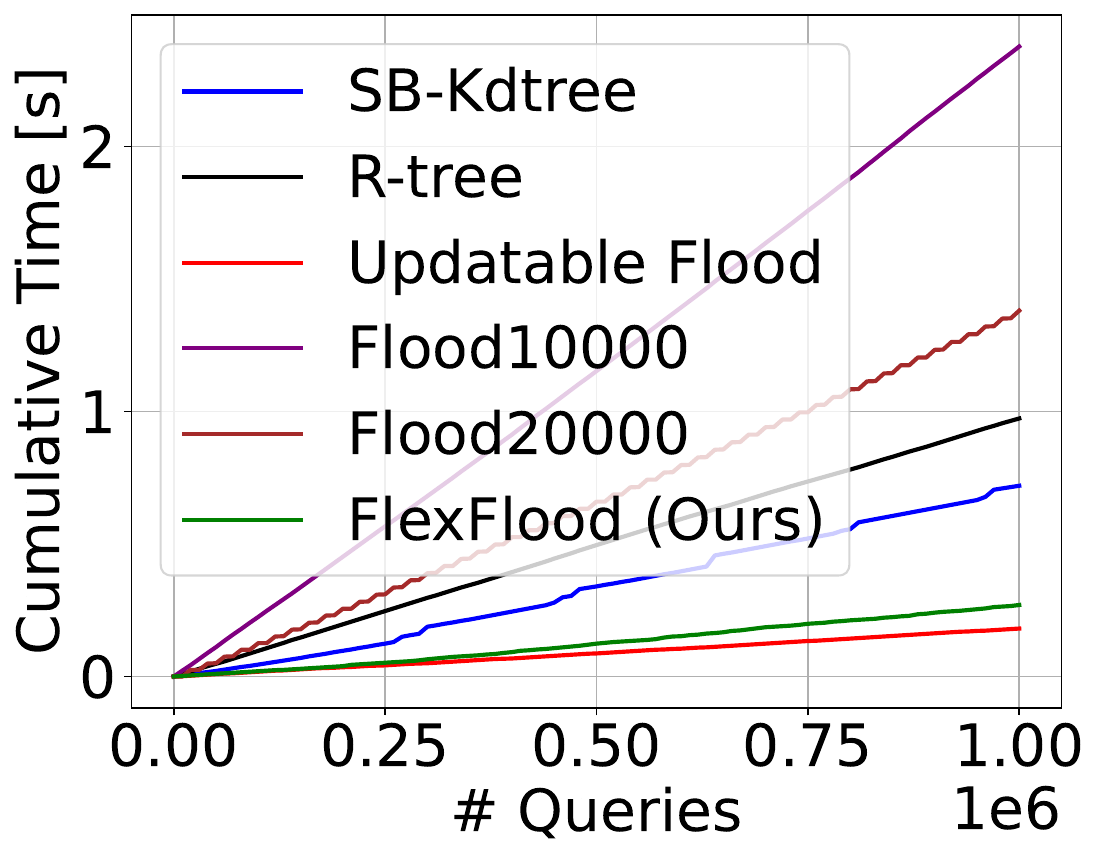}
        \caption{Normal Distribution Update}
        \label{Normal3DUpdateReInit}
    \end{subfigure}
    \hfill
    \begin{subfigure}[tb]{0.32\textwidth}
        \centering
        \includegraphics[width=\textwidth]{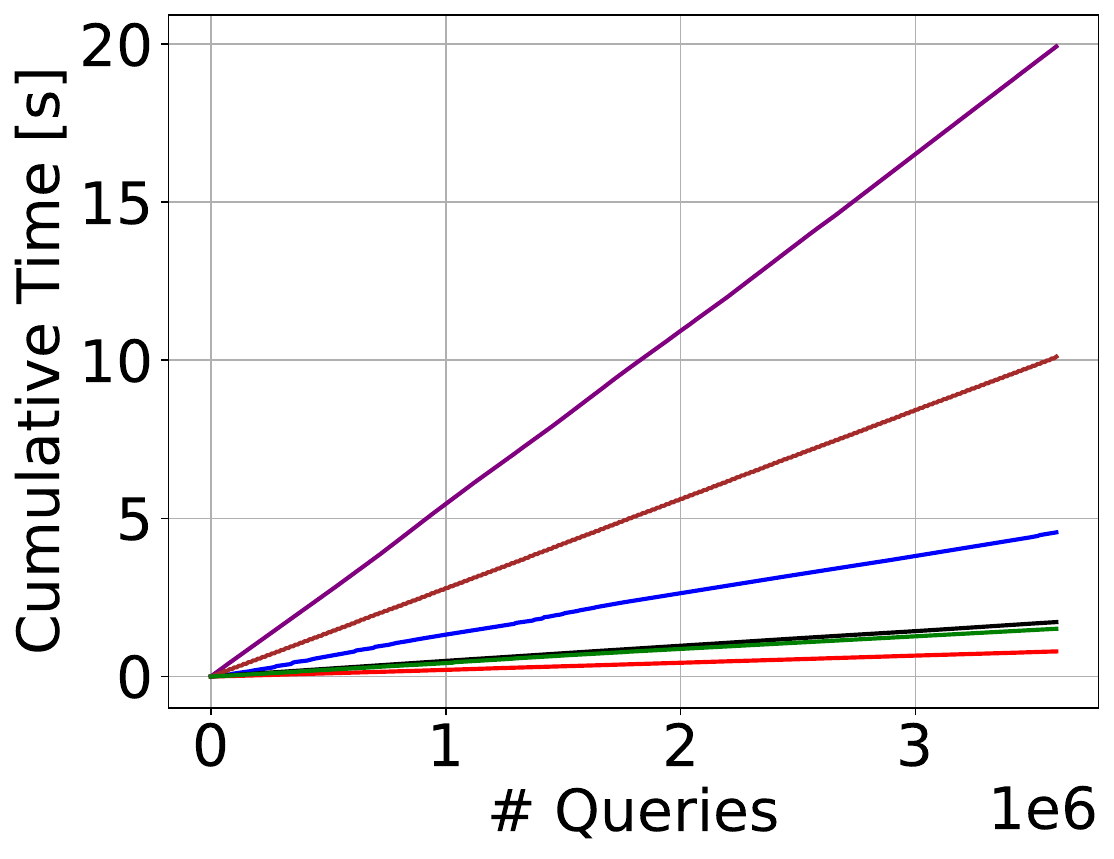}
        \caption{Stock Price Update}
        \label{Stock4DUpdateReInit}
    \end{subfigure}
    \hfill
    \begin{subfigure}[tb]{0.32\textwidth}
        \centering
        \includegraphics[width=\textwidth]{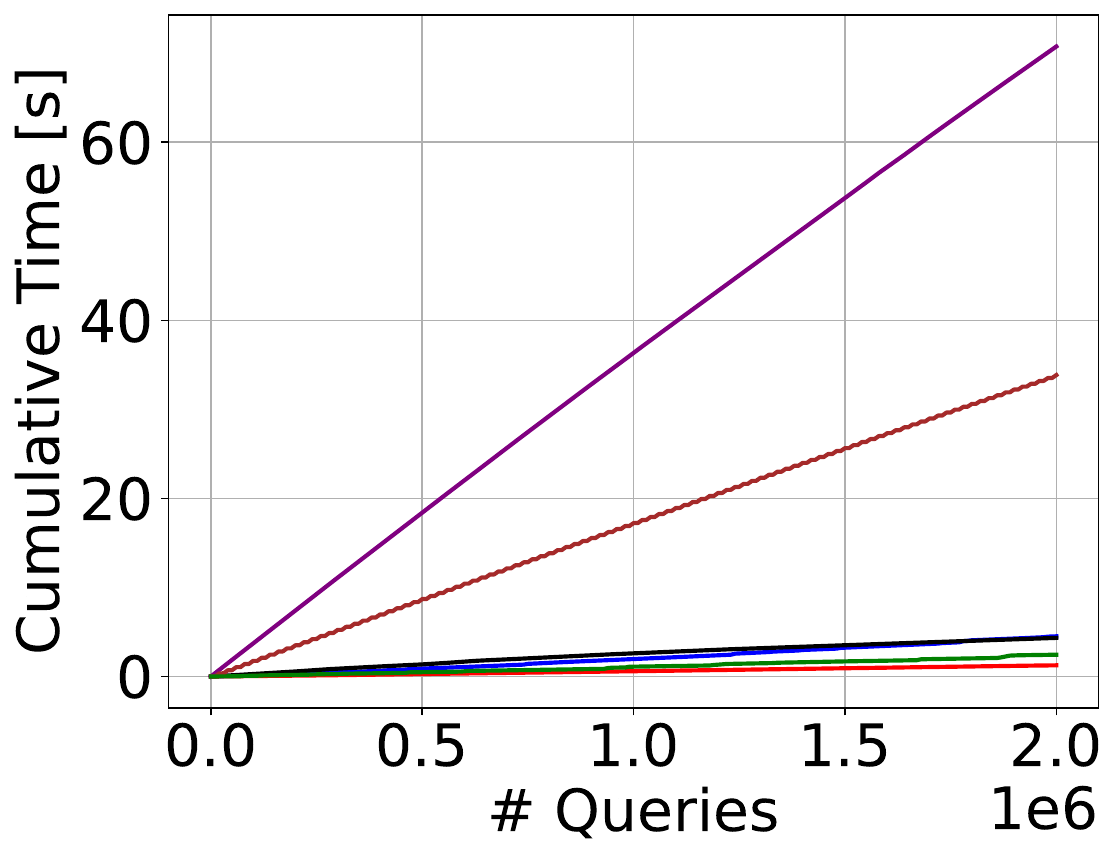}
        \caption{Open Street Map Update}
        \label{OSM5DUpdateReInit}
    \end{subfigure}
    \\
    \begin{subfigure}[tb]{0.32\textwidth}
        \centering
        \includegraphics[width=\textwidth]{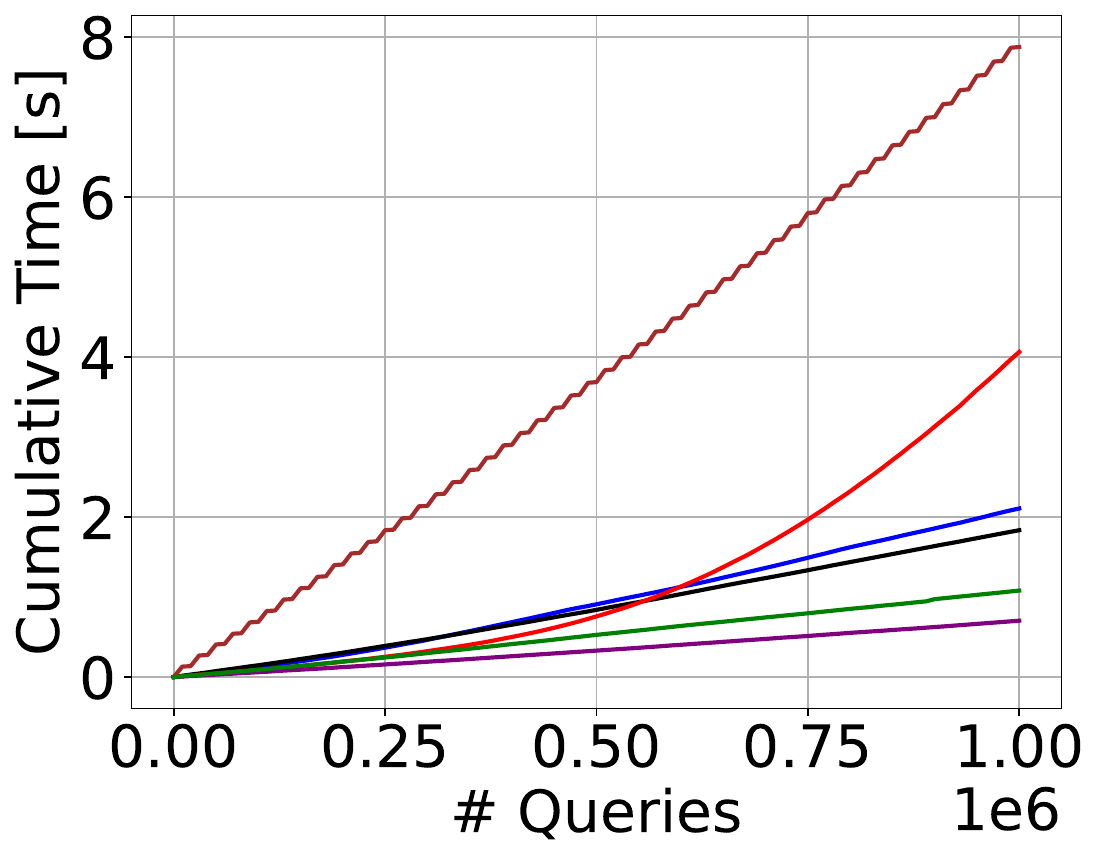}
        \caption{Normal Distribution Search}
        \label{Normal3DReadReInit}
    \end{subfigure}
    \hfill
    \begin{subfigure}[tb]{0.32\textwidth}
        \centering
        \includegraphics[width=\textwidth]{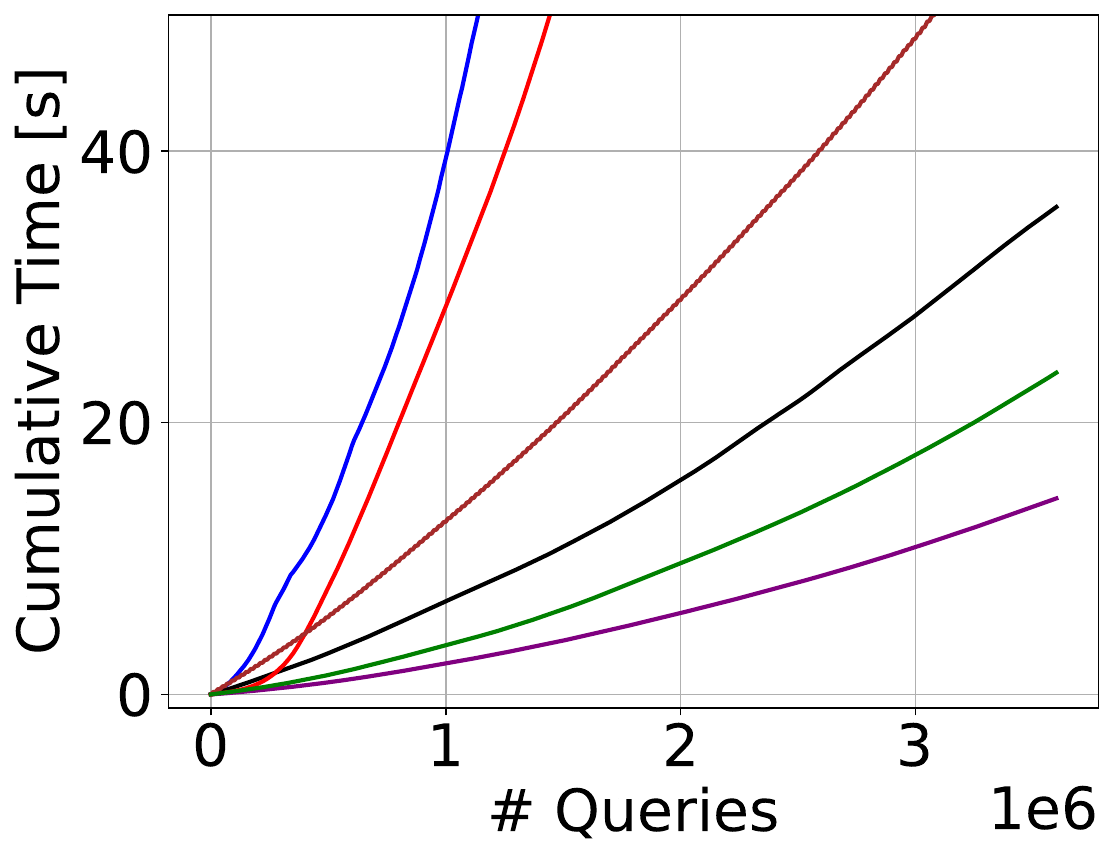}
        \caption{Stock Price Search}
        \label{Stock4DReadReInit}
    \end{subfigure}
    \hfill
    \begin{subfigure}[tb]{0.32\textwidth}
        \centering
        \includegraphics[width=\textwidth]{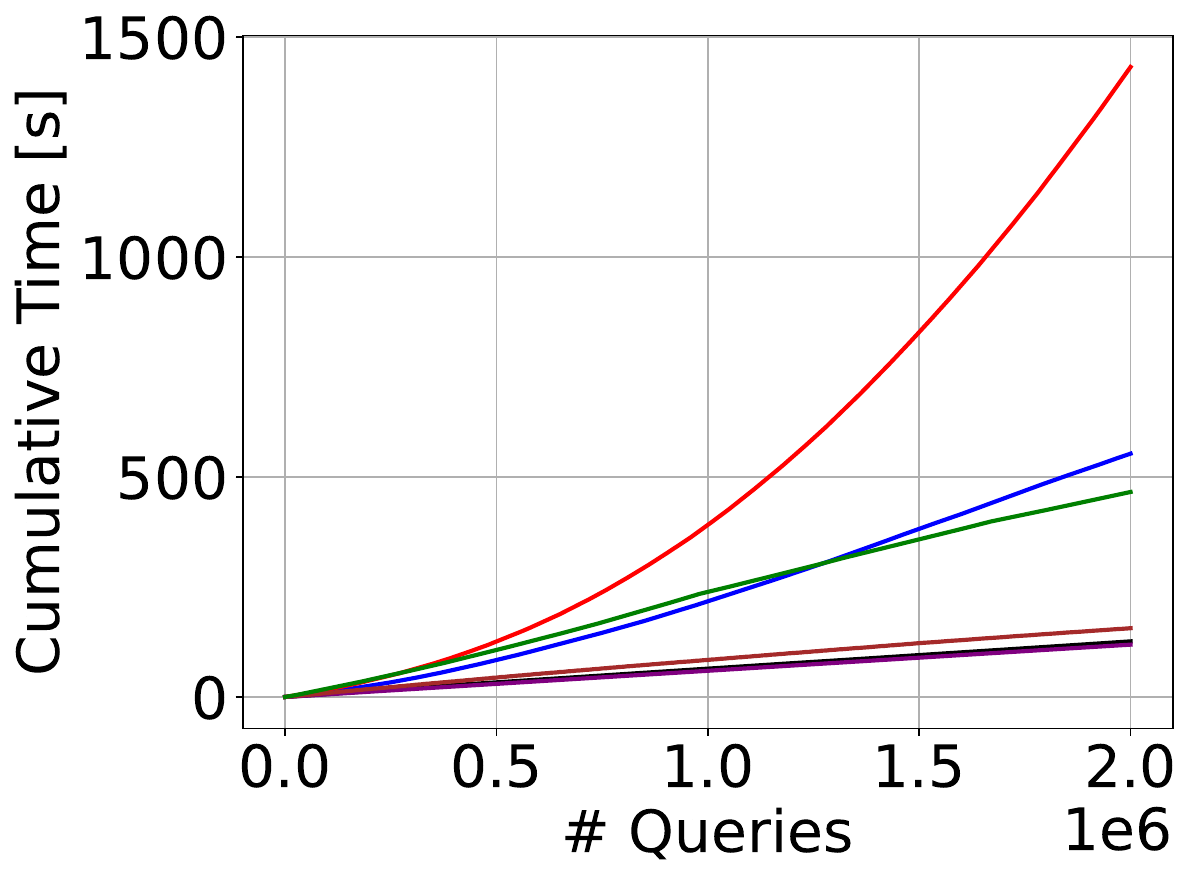}
        \caption{Open Street Map Search}
        \label{OSM5DReadReInit}
    \end{subfigure}
\caption{Experimental results: The upper panel shows the update queries, and the lower panel shows the results for the search queries. (Lower is better.)} 
\label{fig:ResultReInit}
\end{figure*}

Figures~\ref{fig:ResultReInit} illustrate the cumulative query processing time for each data structure for each dataset. Looking at the update query results (Figures~\ref{Normal3DUpdateReInit}, ~\ref{Stock4DUpdateReInit}, ~\ref{OSM5DUpdateReInit}), Flood10000 and Flood20000 are slower than all the compared data structures. Especially, both are more than five times slower than FlexFlood. From this result, the delta buffer Flood is undesirable when there are many update queries. 

We then turn our attention to the search query results (Figures~\ref{Normal3DReadReInit}, ~\ref{Stock4DReadReInit}, ~\ref{OSM5DReadReInit}). Flood10000 has the fastest search speed of all data structures. This is because Flood10000 re-initializes every update query set, so Flood10000 can achieve virtually the same performance as Flood, and the original Flood is extremely fast, as discussed in Appendix~\ref{sec:SortedArrayvsBTree}. On the other hand, Flood20000 is slower than FlexFlood for normal distribution dataset and Stock Prices dataset, but achieves a very fast search speed for Open Street Map dataset. This is likely due to the large amount of data being searched in the Open Street Map dataset (See Appendix~\ref{sec:DatasetDetails}). The time spent searching the internal structure of the Flood has become the rate-limiting step, making the time spent reading from the buffer relatively negligible.

Therefore, delta-buffer Flood is worth considering in situations where an extremely large amount of data is being searched, or when spending a relatively long time on updates is not an issue. On the other hand, we believe that FlexFlood excels in scenarios where high update speed is required or when the amount of data being searched is relatively small.

\end{document}